\documentclass{ifacconf}

\usepackage{graphicx}      
\usepackage{natbib}        

\usepackage{enumerate}
\usepackage{mathtools}
\usepackage{amsmath} 
\usepackage{amssymb}  
\usepackage{etoolbox}
\AtEndEnvironment{pf}{\null\hfill$\square$}%
\usepackage{xcolor}
\usepackage{url}
\usepackage{microtype} 

\usepackage{abbreviations}
\usepackage{tikz}
\usetikzlibrary{shapes,arrows,positioning,fit,backgrounds,calc,plotmarks}
\usepackage{pgfplots}

\usepackage{siunitx}
\usepackage{aligned-overset}

\renewenvironment{thebibliography}[1]{%
    \begin{oldthebibliography}{#1}%
    \setlength{\parskip}{0ex}%
    \setlength{\itemsep}{0ex}%
}%
{%
    \end{oldthebibliography}%
}

\begin{document} 
\setlength{\abovedisplayshortskip}{0.5ex plus1ex minus1ex}
\setlength{\abovedisplayskip}{0.5ex plus1ex minus1ex}
\setlength{\belowdisplayshortskip}{1ex plus1ex minus1ex}
\setlength{\belowdisplayskip}{1ex plus1ex minus1ex}
\begin{frontmatter}

\title{Robust data-driven control for nonlinear systems using the Koopman operator\thanksref{footnoteinfo}} 

\thanks[footnoteinfo]{F.\ Allgöwer is thankful that this work was funded by the Deutsche Forschungsgemeinschaft (DFG, German Research Foundation) under Germany's Excellence Strategy -- EXC 2075 -- 390740016 and within grant AL 316/15-1 -- 468094890. 
R.\ Strässer thanks the Graduate Academy of the SC SimTech for its support.\\
© 2023 the authors. This work has been accepted to IFAC for publication under a Creative Commons Licence CC-BY-NC-ND.}

\author{Robin Strässer,} 
\author{Julian Berberich,} 
\author{Frank Allgöwer} 

\address{\hspace*{-1cm}University of Stuttgart, Institute for Systems Theory and Automatic Control,\hspace*{-0.8cm}~\\Germany (e-mail: \{straesser,berberich,allgower\}@ist.uni-stuttgart.de).}

\begin{abstract}                
Data-driven analysis and control of dynamical systems have gained a lot of interest in recent years. 
While the class of linear systems is well studied, theoretical results for nonlinear systems are still rare. 
In this paper, we present a data-driven controller design method for discrete-time control-affine nonlinear systems.  
Our approach relies on the Koopman operator, which is a linear but infinite-dimensional operator lifting the nonlinear system to a higher-dimensional space.
Particularly, we derive a linear fractional representation of a lifted bilinear system representation based on measured data.
Further, we restrict the lifting to finite dimensions, but account for the truncation error using a finite-gain argument. 
We derive a linear matrix inequality based design procedure to guarantee robust local stability for the resulting bilinear system for all error terms satisfying the finite-gain bound and, thus, also for the underlying nonlinear system. 
Finally, we apply the developed design method to the nonlinear Van der Pol oscillator.
\end{abstract}


\end{frontmatter}

\section{Introduction}\label{sec:introduction}
Designing stabilizing controllers for nonlinear systems is in general a challenging task in engineering and control theory. To apply established nonlinear controller design techniques, typically an accurate nonlinear system model is required~\citep{khalil:2002}. However, obtaining such a model via first-principles or system identification is often time-consuming and requires expert knowledge. Therefore, an alternative strategy gained increasing interest in the last decade, where data-driven methods are applied for controller design~\citep{hou:wang:2013}.

The fundamental characterization of linear time-invariant (LTI) systems via measured data given by~\cite{willems:rapisarda:markovsky:demoor:2005} provided a promising framework for linear data-driven controller design. 
Within this framework,~\cite{depersis:tesi:2019} design state- and output-feedback controllers based on noise-free data. 
The results in~\cite{berberich:koch:scherer:allgower:2020,waarde:camlibel:mesbahi:2020} extend these findings by designing robust controllers based on noisy data, and a framework to additionally include prior knowledge in the data-driven controller design is formulated in~\cite{berberich:scherer:allgower:2022}. 
However, most approaches in the literature focus on LTI systems such that the results are not easily applicable to nonlinear systems. 
Possible extensions have been developed, e.g., 
for Hammerstein and Wiener systems~\citep{berberich:allgower:2020}, 
bilinear systems~\citep{bisoffi:depersis:tesi:2020a}, 
polynomial systems~\citep{guo:depersis:tesi:2021,dai:sznaier:2021}, 
and rational dynamics~\citep{strasser:berberich:allgower:2021}. 
The latter work has also introduced a controller design method for nonlinear systems which can be linearly factorized into non-polynomial basis functions. 
Though, closed-loop stability is still an open problem for more general classes of nonlinear systems.

A promising approach to analyze and control nonlinear systems relies on the Koopman operator~\citep{koopman:1931}, where the nonlinear dynamics are first lifted to a (possibly) infinite-dimensional linear system and then the controller can be obtained based on linear design methods for a finite-dimensional approximation. 
Almost 75 years after \citeauthor{koopman:1931}'s seminal work in \citeyear{koopman:1931}, \cite{mezic:2005} revive the Koopman operator and propose its use for prediction and control. 
In the following decade, a number of contributions have been made, including 
the analysis of global stability properties~\citep{mauroy:mezic:2016}, 
estimation~\citep{netto:mili:2018}, 
and numerical methods for approximating the Koopman action based on data, such as extended dynamic mode decomposition (EDMD)~\citep{williams:kevrekidis:rowley:2015} and its analysis~\citep{korda:mezic:2018b,haseli:cortes:2022}. 
While the Koopman theory was originally developed for autonomous systems, the literature also contains extensions to controlled systems, 
e.g., model predictive control~\citep{korda:mezic:2018a}, 
linear-quadratic regulation~\citep{brunton:brunton:proctor:kutz:2016}, 
and~\cite{lian:wang:jones:2021} provide a Koopman perspective on the work by~\cite{willems:rapisarda:markovsky:demoor:2005}. 
Although Koopman control was shown to work well in numerous applications~\citep{budisic:mohr:mezic:2012}, e.g., robotics~\citep{bruder:remy:vasudevan:2019} or power grids~\citep{korda:susuki:mezic:2018}, deriving error bounds is an active field of research~\citep{nuske:peitz:philipp:schaller:worthmann:2021}. 
Thus, the Koopman-based control methods often lack closed-loop stability guarantees.

In this paper, we propose a data-driven controller design method with stability guarantees for discrete-time control-affine nonlinear systems. 
Inspired by Koopman operator theory, we use a lifted bilinear representation of the underlying system based on measured data.
Moreover, we consider a finite-dimensional approximation, but account for the truncation error using a finite-gain argument.
Then, we obtain a linear matrix inequality (LMI) based design procedure guaranteeing robust local stability for the resulting bilinear system for all errors consistent with the finite-gain bound. 
Thus, the proposed controller also stabilizes the underlying nonlinear system.
This approach is inspired by the work in~\cite{sinha:nandanoori:drgona:vrabie:2022,mamakoukas:dicairano:vinod:2022}. 
The authors of the first reference derive an LMI to find a local stabilizing controller for a \emph{nominal} bilinear system representing the finite-dimensional approximation of a Koopman lifting. 
Since they neglect the truncation error, the obtained controller is not guaranteed to stabilize the closed loop of the true nonlinear system. 
Moreover,~\cite{mamakoukas:dicairano:vinod:2022} derive a \emph{linear} Koopman representation of a controlled nonlinear system while characterizing the truncation error via Lipschitz-continuity to bound the prediction error in a model predictive control framework. 
However, the \emph{bi}linear representation is more accurate and, hence, leads to a less conservative controller design in comparison to a linear one.

\emph{Outline:}
The paper is organized as follows. 
In Section~\ref{sec:Koopman}, we introduce the Koopman operator and its action on autonomous and controlled systems. 
The Koopman reformulation of a nonlinear system as a bilinear system with finite-gain bounded truncation error is presented in Section~\ref{sec:bilinear-representation}.
Based on this, Section~\ref{sec:controller} contains the main theoretical result: a controller design method with guaranteed stability based on robust control arguments. 
In Section~\ref{sec:numerical-example}, the developed approach is applied to a Van der Pol oscillator in simulation. 
Finally, we conclude the paper in Section~\ref{sec:conclusion}.

\emph{Notation:}
We write $I_p$ for the $p\times p$ identity matrix and $0_{p\times q}$ for the $p\times q$ zero matrix, where we omit the index if the dimension is clear from the context. 
If $A$ is symmetric, we write $A\succ 0$ or $A\succeq 0$ if $A$ is positive definite or positive semidefinite, respectively. 
Negative (semi)definiteness is defined analogously. 
We denote matrix blocks which can be inferred from symmetry by $\star$ and abbreviate $B^\top A B$ by writing $[\star]^\top AB$. 
Finally, $\kron$ denotes the Kronecker product.

\section{Koopman Operator}\label{sec:Koopman}
In this section, we give a brief background on Koopman operator theory for autonomous systems and its extension to controlled systems.

\emph{Autonomous systems:}
Consider a nonlinear dynamical system defined by 
$
    x_{k+1} = f(x_k)
$
with state space $x\in\cX\subseteq\bbR^n$, nonlinear state transition map $f: \cX \to \cX$, and discrete time index $k\geq 0$.
In Koopman theory, scalar functions are used to characterize the evolution of the system, commonly referred to as \emph{observables}, by studying the action of the Koopman operator $\cK: \cF \to \cF$,
$
    \cK \phi = \phi \circ f
$,
for every observable $\phi:\cX \to \bbC$, $\phi\in\cF$. Here, $\cF$ is a space of functions invariant under the action of the Koopman operator and $\circ$ denotes function composition.
For an arbitrary state $x_k$, the Koopman action is equivalent to 
$
    \cK \phi(x_k) = \phi \circ f(x_k) = \phi(x_{k+1})
$.
Although the underlying dynamical system is nonlinear and finite-dimensional, the Koopman operator is \emph{linear} but typically \emph{infinite}-dimensional.
Unlike a linearization based on a first-order Taylor expansion, the linear Koopman operator describes the nonlinear system \emph{globally}.

Of particular interest is the truncation error between the (infinite-dimensional) Koopman operator and its approximation resulting by the restriction to a \emph{finite-dimensional} space of observables $\cF_\cD\subseteq \cF$. Here, $\cD =\{\phi_j\}_{j=1}^{N}$ is called \emph{dictionary}. 
If $\cF_\cD$ is invariant under the action of the Koopman operator and rich enough to describe all nonlinearities of the underlying system, the approximation is exact and no error occurs.
Otherwise, the choice of the dictionary $\cD$ influences the size of the resulting truncation error.
Moreover, the finite-dimensional lifted Koopman system reads $z_{k+1} = A z_k$, where $z=\Phi(x)\in\bbC^{N}$ and $\Phi(x) = \begin{bmatrix}
    \phi_1(x) & \cdots & \phi_N(x)
\end{bmatrix}^\top$.

\emph{Controlled systems:}
Now, we consider a nonlinear discrete-time system with control-affine dynamics of the form 
\begin{equation}\label{eq:dynamics-nonlinear}
    x_{k+1} = f(x_k) + g(x_k) u_k
\end{equation}
with state space $x\in\cX\subseteq \bbR^n$, scalar input $u\in\bbR$, and continuously differentiable nonlinear maps $f,g:\cX\to\bbR^n$.
While early works assume a linear representation of the Koopman lifting of controlled systems~\citep{korda:mezic:2018a}, a more accurate lifting is obtained by generalizing the Koopman framework to a bilinear representation. 
A switched-system approach leads in continuous-time to an exact bilinear formulation
~\citep[see, e.g.,][Ch.~4,~12]{surana:2016,mauroy:mezic:susuki:2020} using the Koopman operator. 
Following this observation, we use a finite-dimensional lifted bilinear representation of the controlled system
\begin{align}\label{eq:lifting-dynamics-controlled-nonlinear}
    z_{k+1} &= \Phi(x_{k+1}) = \Phi(f(x_k) + g(x_k)u_k) 
    \\\nonumber
    &= \Phi(f(x_k)) + u_k\left(B_0 + \frac{\partial \Phi}{\partial x}g(x_k)\right) + \tilde{\varepsilon}(z_k,u_k),
\end{align}
where $\tilde{\varepsilon}$ denotes the error resulting from neglecting higher-order terms. Moreover, we write $\Phi(f(x_k))=Az_k+\varepsilon_f(z_k,u_k)$ and $\frac{\partial \Phi}{\partial x}g(x_k)=B_1z_k+\varepsilon_g(z_k,u_k)$, where $A,B_1\in\bbR^{N\times N}$. A common assumption in available literature is that $z=\Phi(x)$ is chosen such that $\varepsilon_f = \varepsilon_g=0$. However, the error is non-zero for general lifting functions $\Phi$. Hence, we express~\eqref{eq:lifting-dynamics-controlled-nonlinear} as the error-affected bilinear control system
\begin{equation}\label{eq:dynamics-bilinear}
    z_{k+1} = A z_k + u_k(B_0 + B_1 z_k) + \varepsilon(z_k,u_k),
\end{equation}
where 
$
    \varepsilon(z_k,u_k) = \tilde{\varepsilon}(z_k,u_k) + \varepsilon_f(z_k,u_k) + u_k\varepsilon_g(z_k,u_k)
$
denotes the truncation error.
For simplicity, we omit the arguments of $\varepsilon(z_k,u_k)$ and write $\varepsilon_k$. 

\section{Bilinear representation of nonlinear systems}\label{sec:bilinear-representation}
\vspace{-0.2cm}
In the remainder of the paper, we use the bilinear representation~\eqref{eq:dynamics-bilinear} as a basis to design stabilizing controllers $u=k^\top z$, $k\in\bbR^{N}$, for the nonlinear system~\eqref{eq:dynamics-nonlinear}. To this end, we first derive the finite-dimensional bilinear representation in Section~\ref{sec:problem-setting}. In Section~\ref{sec:EDMD}, we then use EDMD to obtain the unknown system matrices from data and address the truncation error of the finite-dimensional approximation of the Koopman operator according to Section~\ref{sec:Koopman}.

\vspace{-0.1cm}
\subsection{Problem setting}\label{sec:problem-setting}
\vspace{-0.2cm}
Throughout the paper, we consider nonlinear systems as in~\eqref{eq:dynamics-nonlinear} with an equilibrium at $(x,u)=0$, i.e., $f(0)=0$. 
Lifting the system dynamics to a higher-dimensional state $z=\Phi(x)$ leads to the bilinear representation~\eqref{eq:dynamics-bilinear}, where $A$, $B_0$, $B_1$, and $\varepsilon_k$ are unknown. 
Since we also account for the truncation error $\varepsilon_k$, the bilinear representation equivalently describes the nonlinear dynamics~\eqref{eq:dynamics-nonlinear}. 
Thus, we design a control law $u=k^\top z$ robustly stabilizing~\eqref{eq:dynamics-bilinear} locally for all $z$ in an ellipsoidal region $\cZ_\mathrm{RoA}$ defined later, and show that this controller also locally stabilizes the nonlinear system. Moreover, we define the set 
\begin{equation*}
    \cZ = \left\{
        z \in \bbR^N
    \middle|
        \begin{bmatrix}
            z \\ 1
        \end{bmatrix}^\top 
        \begin{bmatrix}
            Q_z & S_z \\ S_z^\top & R_z
        \end{bmatrix}
        \begin{bmatrix}
            z \\ 1
        \end{bmatrix} \geq 0
    \right\}
\end{equation*}
with $Q_z\prec 0$ and $R_z\succ 0$. This includes, e.g., the bound $z^\top z \leq c$ for $Q_z=-I$, $S_z = 0$, and $R_z = c$, $c>0$. 
Further, we assume that the inverse
\begin{equation*}
    \begin{bmatrix}
        Q_z & S_z \\ S_z^\top & R_z
    \end{bmatrix}^{-1}
    = \begin{bmatrix}
        \tQ_z & \tS_z \\ \tS_z^\top & \tR_z
    \end{bmatrix}
\end{equation*}
exists.
For the theoretical analysis, we exploit that the lifted state $z$ is within $\cZ$ for all times which is ensured via a suitable Lyapunov function sublevel set $\cZ_\mathrm{RoA}\subseteq \cZ$.

\begin{assum}\label{ass:zero-if-zero}
    The lifting $\Phi$ satisfies $\Phi(x) = 0$ iff $x=0$.
\end{assum}
This assumption ensures that the bilinear system has an equilibrium at $(z,u)=0$.  
As described in Section~\ref{sec:Koopman}, the choice of a suitable lifting function $\Phi$ is crucial for a small truncation error and therefore an accurate bilinear representation. How to select this function for a general nonlinear system is still an open research question~\citep{iacob:toth:schoukens:2022}, where input-dependent liftings $\Psi(x,u)$ are proposed instead of solely state-dependent liftings $\Phi(x)$. 

\emph{Delay coordinates:}
One common choice of the lifting is via so called delay coordinates~\citep[cf.][Chapter 8]{mamakoukas:dicairano:vinod:2022,mauroy:mezic:susuki:2020}, i.e., 
$
    z_k = \Psi(x_{k-d_x:k},u_{k-d_u:k})
$\\
$= \begin{bmatrix}
        x_k^\top & h_k^\top
    \end{bmatrix}^\top
$
with $z\in\bbR^{N}$ and the history of states and inputs $
    h_k = \begin{bmatrix}
         h_{x,k}^\top & h_{u,k} & h_{xu,k}^\top
    \end{bmatrix}^\top
$.
Besides the history of the state 
$
    h_{x,k} = \begin{bmatrix}
        x_{k-1}^\top & \cdots & x_{k-d_x}^\top
    \end{bmatrix}^\top
$ and input $
    h_{u,k} = \begin{bmatrix}
        u_{k-1} & \cdots & u_{k-d_u}
    \end{bmatrix}^\top
$, we also store the history of its product $
    h_{xu,k} = \begin{bmatrix}
        x_{k-1}^\top u_{k-1} & \cdots & x_{k-d_u}^\top u_{k-d_u}
    \end{bmatrix}^\top
$ in the proposed lifting to obtain a proper delay structure of the bilinear system~\eqref{eq:dynamics-bilinear}.
The numbers of delay states and delay control inputs are denoted by $d_x$ and $d_u$, respectively, where $d_x\geq d_u$, and, thus, the lifting dimension is $N=n(1+d_x+d_u) + d_u$. 
Moreover, the true state $x$ can be exactly recovered via the linear equation $x_k=\begin{bmatrix}I&0\end{bmatrix} z_k$.
The dynamics based on this delay lifting read 
\newlength\mylen
\settoheight\mylen{$
    z_{k+1}
    = \begin{bmatrix}
        A_\mathrm{u} \\ A_\mathrm{k}
    \end{bmatrix}
    \begin{bmatrix}
        x_k \\ h_k
    \end{bmatrix}
    + u_k \left(
        \begin{bmatrix}
            B_{0,\mathrm{u}} \\ B_{0,\mathrm{k}}
        \end{bmatrix}
        + 
        \begin{bmatrix}
            B_{1,\mathrm{u}} \\ B_{1,\mathrm{k}}
        \end{bmatrix}
        \begin{bmatrix}
            x_k \\ h_k
        \end{bmatrix}
    \right)
    + \begin{bmatrix}
        \varepsilon_{x,k} \\ 0
    \end{bmatrix}
$}
\begin{equation}\label{eq:dynamics-bilinear-structured}
\hspace*{-5.5pt}
\resizebox{0.93\linewidth}{\mylen}{$
    z_{k+1}
    = \begin{bmatrix}
        A_\mathrm{u} \\ A_\mathrm{k}
    \end{bmatrix}
    \begin{bmatrix}
        x_k \\ h_k
    \end{bmatrix}
    + u_k \left(
        \begin{bmatrix}
            B_{0,\mathrm{u}} \\ B_{0,\mathrm{k}}
        \end{bmatrix}
        + 
        \begin{bmatrix}
            B_{1,\mathrm{u}} \\ B_{1,\mathrm{k}}
        \end{bmatrix}
        \begin{bmatrix}
            x_k \\ h_k
        \end{bmatrix}
    \right)
    + \begin{bmatrix}
        \varepsilon_{x,k} \\ 0
    \end{bmatrix}
$}
\end{equation}
with \emph{known} matrices%
{%
    \begin{gather*}
    {
        A_\mathrm{k} = \diag\left( 
            \begin{bmatrix} 
                I_{nd_x} & 0 
            \end{bmatrix},
            \begin{bmatrix}
                0 & 0 \\
                I_{d_u-1} & 0
            \end{bmatrix},
            \begin{bmatrix}
                0 & 0 \\
                I_{n(d_u-1)} & 0
            \end{bmatrix}
        \right), 
    }
    \\
    {
        B_{0,\mathrm{k}} = \begin{bmatrix}
            0_{nd_x\times 1}
            \\
            \begin{bmatrix} 1 \\ 0_{d_u-1\times 1} \end{bmatrix}
            \\
            0_{nd_u\times 1}
        \end{bmatrix},
        \,
        B_{1,\mathrm{k}} = \begin{bmatrix}
            0_{nd_x \times n} & 0_{nd_x \times N-n}\\ 0_{d_u\times n} & 0_{d_u\times N-n} \\ \begin{bmatrix} I_n \\ 0_{n(d_u-1)\times n} \end{bmatrix} & 0_{nd_u\times N-n}
        \end{bmatrix}
    }
    \end{gather*}%
}%
according to the delay structure in the history, and \emph{unknown} matrices $A_\mathrm{u}$, $B_{0,\mathrm{u}}$, $B_{1,\mathrm{u}}$ depending on the system dynamics.
In particular, the resulting error is explicitly defined by the truncation error in the state $x$, i.e.,
\begin{equation*}
    \varepsilon_{x,k} = f(x_k) + g(x_k) u_k 
    - A_\mathrm{u}z_k - u_k(B_{0,\mathrm{u}} + B_{1,\mathrm{u}}z_k).
\end{equation*}
As $\varepsilon_{x,k}$ is directly related to the nonlinear functions $f,g$, a bound on the error $\varepsilon_{x,k}$ and, thus, $\varepsilon_k$ can be derived, e.g., via Lipschitz continuity if $f$ and $g$ are Lipschitz continuous.

\emph{Monomial lifting:}
While the considered delay lifting in the last paragraph results in a structured dynamics, also non-structured lifting functions $\Phi$ are possible leading to fully unknown matrices $A=A_\mathrm{u}$, $B_0=B_{0,\mathrm{u}}$, $B_1=B_{1,\mathrm{u}}$, and an unknown inverse lifting $x_k=\Phi^{-1}(z_k)$ that might introduce an additional error.
A widely used lifting consists of monomials~\citep[cf.][]{carleman:1932}, where the true state is exactly recovered via $x_k=\begin{bmatrix}I&0\end{bmatrix} z_k$. 
Here, every coordinate of $z_k$ is affected by the truncation error $\varepsilon_k$ and thus, it is more difficult to derive non-conservative error bounds.

Other commonly used liftings rely, e.g., on radial basis functions or on an additional nonlinear transformation of delay coordinates.

\vspace{-0.1cm}
\subsection{Extended dynamic mode decomposition}\label{sec:EDMD}
\vspace{-0.1cm}
In the following, we estimate the unknown system matrices $A$, $B_0$, and $B_1$ in~\eqref{eq:dynamics-bilinear} from data using EDMD. To this end, we collect data $\{\{x_k^\mathrm{d}\}_{k=-d_x}^{L},\{u_k^\mathrm{d}\}_{k=-d_x}^{L-1}\}$ and arrange the samples in data matrices
$
    U = \begin{bmatrix}
        u_0^\mathrm{d} & \cdots & x_{L-1}^\mathrm{d}
    \end{bmatrix}
$, $
    Z = \begin{bmatrix}
        \Phi(x_0^\mathrm{d}) & \cdots & \Phi(x_{L-1}^\mathrm{d})
    \end{bmatrix}
$, and $
    Z_+ = \begin{bmatrix}
        \Phi(x_1^\mathrm{d}) & \cdots & \Phi(x_L^\mathrm{d})
    \end{bmatrix}
$. 
Further, we define $Y=\begin{bmatrix}
    Z^\top & U^\top & (Z \diag(U))^\top
\end{bmatrix}^\top$.
Then, we obtain the unknown matrices by the least-squares problem
$ 
    \min_{A,B_0,B_1}
    \|
        Z_+ 
        - \begin{bmatrix}
            A & B_0 & B_1
        \end{bmatrix}
        Y
    \|
$
with the $\ell_2$-optimal solution $
    \begin{bmatrix}
        A & B_0 & B_1
    \end{bmatrix}
    = Z_+ Y^\dagger
$, where $Y^\dagger$ is the pseudoinverse of $Y$. 
This procedure yields optimal system matrices satisfying the dynamics~\eqref{eq:dynamics-bilinear} such that the error $\varepsilon_k$ is minimized.

The optimization can be simplified depending on the lifting structure. For instance, a delay lifting results only in partly unknown system matrices $A_\mathrm{u}$, $B_{0,\mathrm{u}}$, and $B_{1,\mathrm{u}}$ in~\eqref{eq:dynamics-bilinear-structured}. Thus, we define $X_+ = \begin{bmatrix}
    x_1^\mathrm{d} & \cdots & x_L^\mathrm{d}
\end{bmatrix}$ 
and solve the reduced optimization $
    \min_{A_\mathrm{u},B_{0,\mathrm{u}},B_{1,\mathrm{u}}}
    \|
        X_+ 
        - \begin{bmatrix}
            A_\mathrm{u} & B_{0,\mathrm{u}} & B_{1,\mathrm{u}}
        \end{bmatrix}
        Y
    \|
$, where $Y$ is defined as before but with
$Z = \left[\begin{array}{cc}
        \Psi(x_{-d_x:0}^\mathrm{d},u_{-d_u:0}^\mathrm{d}) & \cdots 
    \end{array}\right.
$ $\left.\begin{array}{c}
        \Psi(x_{L-1-d_x:L-1}^\mathrm{d},u_{L-1-d_u:L-1}^\mathrm{d})
    \end{array}\right]
$.

Further, EDMD converges under certain assumptions to the Koopman operator for $N\to\infty$ and $L\to\infty$~\citep{korda:mezic:2018b}, and a probabilistic error analysis is given in~\cite{chen:vaidya:2019,nuske:peitz:philipp:schaller:worthmann:2021}.

Applying now the linear feedback law $u = k^\top z$, the closed-loop dynamics of the bilinear system~\eqref{eq:dynamics-bilinear} read 
\begin{equation*}
    z_{k+1} = (A + B_0k^\top)z_k + B_1 z_k (k^\top z_k) + \varepsilon_k,
\end{equation*}
where $\varepsilon_k=\varepsilon(z_k,k^\top z_k)$ is a function of $z_k$.
\begin{assum}\label{ass:finite-gain}
    The truncation error $\varepsilon_k$ satisfies for all times $k\geq 0$ the finite-gain bound
    \begin{equation*}
        \varepsilon_k \in \cE \coloneqq \left\{
            \varepsilon \in \bbR^N
        \middle|
            \|\varepsilon\| 
            \leq L_\varepsilon 
            \| z \| 
            \>\text{for all}\> z \in \cZ
        \right\}.
    \end{equation*}
\end{assum}

Choosing a suitable lifting function $\Phi$ is crucial for a small constant $L_\varepsilon$ and therefore an accurate bilinear representation. For the remainder of the paper, we assume that the finite-gain bound of the truncation error (or at least an upper bound) is known. Clearly, the approximation accuracy of the bilinear system compared to the true nonlinear system and, thus, the constant $L_\varepsilon$ depend on the chosen lifting. A more rigorous investigation of how to obtain this bound, e.g., by using the approach proposed by~\cite{wood:zhang:1996}, is left for future work.

Moreover, a generalization of the error characterization $\cE$ is an interesting issue for future research, e.g., by considering local Lipschitz continuity w.r.t. each measured data sample instead of just the origin. Particularly, combining knowledge about Lipschitz continuity w.r.t. multiple data points leads to an uncertainty envelope described by an \emph{incremental} quadratic constraint~\citep[cf.][]{calliess:2014}.

\begin{rem}
    Instead of explicitly identifying the unknown matrices from data via EDMD, we can also follow the idea of~\cite{waarde:camlibel:mesbahi:2020,berberich:scherer:allgower:2022} and derive a parametrization of all possible choices of system matrices resulting in a system which is consistent with the data and the assumed envelope of the truncation error. The investigation of such a direct data-driven system parametrization is interesting for future research.
\end{rem} 

\vspace{-0.1cm}
\section{Stabilizing controller design for nonlinear systems}\label{sec:controller}
\vspace{-0.1cm}
Next, we want to design a controller $u = k^\top z$ stabilizing the bilinear system~\eqref{eq:dynamics-bilinear} under consideration of the truncation error and, thus, stabilizing also the underlying \emph{true} nonlinear system~\eqref{eq:dynamics-nonlinear}. In Section~\ref{sec:controller:bilinear}, we first derive a controller design procedure based on LMIs for the nominal bilinear system which we later extend to closed-loop stability of the true nonlinear system in Section~\ref{sec:controller:nonlinear}.

\vspace{-0.1cm}
\subsection{Controller for the nominal bilinear system}\label{sec:controller:bilinear}
\vspace{-0.1cm}
In this section, we provide a procedure to find a local stabilizing controller for the nominal bilinear system 
\begin{equation}\label{eq:dynamics-bilinear-nominal}
    z_{k+1} = A z_k + u_k(B_0 + B_1 z_k)
\end{equation}
for all $z_0\in\cZ_\mathrm{RoA}\subseteq\cZ$.
To apply linear controller design methods, we first rewrite~\eqref{eq:dynamics-bilinear-nominal} as a linear fractional representation (LFR) which is a common description of uncertain systems~\citep{zhou:doyle:glover:1996}. 
In particular, the closed loop with the controller $u=k^\top z$ reads
\begin{equation}\label{eq:LFR-nominal}
        \begin{bmatrix}
            z_{k+1} \\ u_k
        \end{bmatrix}
        = \begin{bmatrix}
            A + B_0k^\top & B_1 \\ k^\top & 0
        \end{bmatrix}
        \begin{bmatrix}
            z_k \\ w_k
        \end{bmatrix}, 
        \quad
        w_k = z_k u_k,
\end{equation} 
where the state $z_k\in\cZ$ is interpreted as an uncertainty. 
The following theorem establishes local closed-loop stability of~\eqref{eq:dynamics-bilinear-nominal} based on~\eqref{eq:LFR-nominal}.

\begin{thm}\label{thm:stability-condition-LFR-nominal}
    Suppose Assumption~\ref{ass:zero-if-zero} holds. If there exist an $N\times N$ matrix ${P}=P^\top\succ 0$ and a vector ${y}\in\bbR^{N}$ such that
    \begin{equation}\label{eq:stability-condition-LFR-nominal}
        \begin{bmatrix}
            {P} + B_1 \tQ_z B_1^\top
            & - B_1 \tS_z
            & A {P} + B_0 {y}^\top
            \\
            -\tS_z^\top B_1^\top
            & \tR_z
            & {y}^\top
            \\
            {P} A^\top + {y} B_0^\top 
            & {y}
            & {P} 
        \end{bmatrix}
        \succ 0,
    \end{equation}
    then the controller $u = {y}^\top {P}^{-1} z$ asymptotically stabilizes system~\eqref{eq:dynamics-bilinear-nominal} for all initial conditions $z_0\in\cZ_\mathrm{RoA}=\{z\in\bbR^N\mid z^\top P^{-1} z \leq c\}$, where $c>0$ such that $\cZ_\mathrm{RoA}\subseteq\cZ$.
\end{thm}
\begin{pf}
    We define $k = P^{-1} y$. Then by using the Schur complement,~\eqref{eq:stability-condition-LFR-nominal} is equivalent to 
    \begin{equation*}
        \begin{bmatrix}
            P +  B_1 \tQ_z B_1^\top
            & -  B_1 \tS_z
            \\
            - \tS_z^\top B_1^\top
            &  \tR_z
        \end{bmatrix}
        - 
        \begin{bmatrix}\star\end{bmatrix}
        P
        \begin{bmatrix}
            (A+B_0 k^\top) \\ k^\top
        \end{bmatrix}^\top
        \succ 0.
    \end{equation*}
    Note that 
    \begin{equation*}
        \begin{bmatrix}
             B_1 \tQ_z B_1^\top
            & - B_1 \tS_z
            \\
            - \tS_z^\top B_1^\top
            &  \tR_z
        \end{bmatrix}
        = \begin{bmatrix}
           \star
        \end{bmatrix}^\top 
        \begin{bmatrix}
            \tQ_z & \tS_z \\ \tS_z^\top & \tR_z
        \end{bmatrix}
        \begin{bmatrix}
            B_1^\top & 0 \\ 0 & -I
        \end{bmatrix}
    \end{equation*}
    and 
    \begin{multline*}
        \begin{bmatrix}
            P & 0 \\ 0 & 0
        \end{bmatrix}
        - 
        \begin{bmatrix}\star\end{bmatrix}
        P
        \begin{bmatrix}
            (A+B_0 k^\top) \\ k^\top
        \end{bmatrix}^\top
        \\
        = \begin{bmatrix}\star\end{bmatrix}^\top
        \begin{bmatrix}
            -P & 0 \\ 0 & P
        \end{bmatrix}
        \begin{bmatrix}
            (A+B_0k^\top)^\top & k \\
            -I & 0
        \end{bmatrix}
    \end{multline*}
    leading to
    \begin{equation*}
        \begin{bmatrix}
            \star
        \end{bmatrix}^\top
        \left[\def\arraystretch{1.15}\begin{array}{cc|cc}
            -P & 0 & 0 & 0 \\ 
            0 & P & 0 & 0 \\\hline
            0 & 0 & \tQ_z &  \tS_z \\
            0 & 0 & \tS_z^\top &  \tR_z
        \end{array}\right]
        \left[\def\arraystretch{1.15}\begin{array}{cc}
            (A+B_0k^\top)^\top & k \\ -I & 0 \\\hline
            B_1^\top & 0 \\ 0 & -I
        \end{array}\right]
        \succ 0.
    \end{equation*}
    Using the dualization lemma~\citep[Lm.~4.9]{scherer:weiland:2000} we obtain 
    \begin{equation*}
        \begin{bmatrix}
            \star
        \end{bmatrix}^\top
        \left[\def\arraystretch{1.15}\begin{array}{cc|cc}
            -P^{-1} & 0 & 0 & 0 \\
            0 & P^{-1} & 0 & 0 \\\hline
            0 & 0 & Q_z & S_z \\
            0 & 0 & S_z^\top & R_z
        \end{array}\right]
        \left[\def\arraystretch{1.15}\begin{array}{cc}
            I & 0 \\
            A+B_0k^\top & B_1 \\\hline
            0 & I \\
            k^\top & 0
        \end{array}\right]
        \prec 0.
    \end{equation*}
    Further, we multiply from left and right by $\begin{bmatrix}z_k^\top & w_k^\top\end{bmatrix}^\top\neq 0$ and its transpose, respectively, and define the Lyapunov function $V(z)=z^\top P^{-1}z$. 
    Then, we conclude $V(z_{k+1})-V(z_k) < 0$ for all $z_k\in\cZ$ due to~\citep[Thm.~10.2]{scherer:2000}. 
    Positive invariance of $\cZ_\mathrm{RoA}$ is guaranteed since $\cZ_\mathrm{RoA}\subseteq\cZ$ is a sublevel set of $V$, i.e., $z_{k+1}\in\cZ_\mathrm{RoA}$ if $z_k\in\cZ_\mathrm{RoA}$ for all $k\geq 0$, which concludes the proof.
\end{pf}

\begin{rem}\label{rem:Sinha-results-nominal}
    \cite{sinha:nandanoori:drgona:vrabie:2022} provide an alternative approach by deriving a local stabilizing controller for system~\eqref{eq:dynamics-bilinear-nominal} based on Petersen's lemma \citep[cf.][]{petersen:1987}. 
    However, their design infers only closed-loop stability if the system matrix $A$ is already stable, whereas Theorem~\ref{thm:stability-condition-LFR-nominal} is also applicable to open-loop unstable systems.
    Moreover, the formulation of Theorem~\ref{thm:stability-condition-LFR-nominal} in the robust control framework of~\cite{scherer:2000} is beneficial for future generalizations of the design procedure.
\end{rem}

Our goal is to guarantee closed-loop stability for the largest region $\cZ_\mathrm{RoA}\subseteq\cZ$. To this end, we first maximize $\tr(P)$ such that~\eqref{eq:stability-condition-LFR-nominal} holds, and second, for a given solution to the design in Theorem~\ref{thm:stability-condition-LFR-nominal}, we maximize $c>0$ such that $\cZ_\mathrm{RoA}\subseteq\cZ$. Further, if there is no predefined region $\cZ$ given, we can also optimize for a large $\cZ$ in the design by, e.g., minimizing $\tr(\tQ_z)$, $\tQ_z\prec 0$ when choosing $S_z=0$, $R_z=1$.

\vspace{-0.1cm}
\subsection{Controller for the nonlinear system}\label{sec:controller:nonlinear}
\vspace{-0.15cm}
Now, we extend the results of the previous section to the nonlinear system~\eqref{eq:dynamics-nonlinear}. To this end, we want to find a controller based on the bilinear representation~\eqref{eq:dynamics-bilinear} while accounting for the truncation error $\varepsilon_k\in\cE$ via the finite-gain bound. Hence, if we find a controller robustly stabilizing system~\eqref{eq:dynamics-bilinear} for all $z\in\cZ_\mathrm{RoA}$ and $\varepsilon\in\cE$, then the controller also locally stabilizes the nonlinear system~\eqref{eq:dynamics-nonlinear}.

First, we write the error-affected bilinear system as an LFR, i.e., the closed loop of~\eqref{eq:dynamics-bilinear} with $u=k^\top z$ reads
\begin{equation}\label{eq:LFR}
        \begin{bmatrix}
            z_{k+1} \\ u_k
        \end{bmatrix}
        = \begin{bmatrix}
            A + B_0k^\top & B_1 & I \\ k^\top & 0 & 0
        \end{bmatrix}
        \begin{bmatrix}
            z_k \\ w_k \\ \varepsilon_k
        \end{bmatrix},
        \quad
        w_k = z_k u_k
\end{equation}
with $z_k\in\cZ$. Now we can state our main theorem.

\begin{thm}\label{thm:stability-condition-LFR}
    Suppose Assumptions~\ref{ass:zero-if-zero} and~\ref{ass:finite-gain} hold. If there exist an $N\times N$ matrix ${P}=P^\top\succ 0$, a vector ${y}\in\bbR^{N}$, and a scalar $\tau > 0$ such that
    \begin{equation}\label{eq:stability-condition-LFR}
        \begin{bmatrix}
            P + B_1\tQ_z B_1^\top - \tau I & -B_1 \tS_z & 0 & AP + B_0 y^\top \\
            -\tS_z^\top B_1^\top & \tR_z & 0 & y^\top \\
            0 & 0 & \tau I & L_\varepsilon P \\
            P A^\top + y B_0^\top & y & L_\varepsilon P & P
        \end{bmatrix}
        \succ 0,
    \end{equation}
    then the controller $u={y}^\top{P}^{-1}z$ asymptotically stabilizes system~\eqref{eq:dynamics-bilinear} for all initial conditions $z_0\in\cZ_\mathrm{RoA}=\{z\in\bbR^N\mid z^\top P^{-1} z\leq c\}$, where $c>0$ such that $\cZ_\mathrm{RoA}\subseteq\cZ$.
\end{thm}
\begin{figure*}[!t] 
    \begin{equation}\label{eq:proof-condition-Schur-complement}
    {\small
        \left[\def\arraystretch{1.15}\begin{array}{ccc}
            (A+B_0y^\top P^{-1})^\top & P^{-1}y & L_\varepsilon I \\
            -I & 0 & 0 \\\hline
            B_1^\top & 0 & 0 \\
            0 & -I & 0 \\\hline
            0 & 0 & -I \\
            I & 0 & 0
        \end{array}\right]^\top
        \left[\def\arraystretch{1.15}\begin{array}{cc|cc|cc}
            -P & 0 & 0 & 0 & 0 & 0 \\ 
            0 & P & 0 & 0 & 0 & 0 \\\hline
            0 & 0 & \tQ_z & \tS_z & 0 & 0 \\ 
            0 & 0 & \tS_z^\top & \tR_z & 0 & 0 \\\hline
            0 & 0 & 0 & 0 & \tau I & 0 \\ 
            0 & 0 & 0 & 0 & 0 & -\tau I
        \end{array}\right]
        \left[\def\arraystretch{1.15}\begin{array}{ccc}
            (A+B_0y^\top P^{-1})^\top & P^{-1}y & L_\varepsilon I \\
            -I & 0 & 0 \\\hline
            B_1^\top & 0 & 0 \\
            0 & -I & 0 \\\hline
            0 & 0 & -I \\
            I & 0 & 0
        \end{array}\right]
        \succ 0
    }
    \end{equation}
    \medskip
    \vspace*{-0.6\baselineskip}
    \hrule
\end{figure*}
\begin{figure*}[!t]
    \vspace*{-0.6\baselineskip} 
    \begin{equation}\label{eq:proof-condition-dualized}
    {\small
        \left[\def\arraystretch{1.15}\begin{array}{ccc}
            I & 0 & 0 \\
            A+B_0k^\top & B_1 & I \\\hline
            0 & I & 0 \\
            k^\top & 0 & 0 \\\hline
            L_\varepsilon I & 0 & 0 \\
            0 & 0 & I
        \end{array}\right]^\top
        \left[\def\arraystretch{1.15}\begin{array}{cc|cc|cc}
            -P^{-1} & 0 & 0 & 0 & 0 & 0 \\ 
            0 & P^{-1} & 0 & 0 & 0 & 0 \\\hline
            0 & 0 & Q_z & S_z & 0 & 0 \\
            0 & 0 & S_z^\top & R_z & 0 & 0 \\\hline
            0 & 0 & 0 & 0 & \tau^{-1} I & 0 \\ 
            0 & 0 & 0 & 0 & 0 & -\tau^{-1} I
        \end{array}\right]
        \left[\def\arraystretch{1.15}\begin{array}{ccc}
            I & 0 & 0 \\
            A+B_0k^\top & B_1 & I \\\hline
            0 & I & 0 \\
            k^\top & 0 & 0 \\\hline
            L_\varepsilon I & 0 & 0 \\
            0 & 0 & I
        \end{array}\right]
        \prec 0
    }
    \end{equation}
    \medskip
    \vspace*{-0.6\baselineskip}
    \hrule
    \vspace*{-0.75\baselineskip}
\end{figure*}
\begin{pf}
    In the following, we show that~\eqref{eq:stability-condition-LFR} implies Lyapunov stability. If~\eqref{eq:stability-condition-LFR} holds, we apply the Schur complement to obtain~\eqref{eq:proof-condition-Schur-complement}.
    Further, we define $k=P^{-1}y$. Then, using the dualization lemma~\citep[Lm.~4.9]{scherer:weiland:2000} results in~\eqref{eq:proof-condition-dualized}.
    Multiplying from left and right by $\begin{bmatrix}
        z_k^\top & w_k^\top & \varepsilon_k^\top
    \end{bmatrix}^\top\neq 0$ and its transpose, respectively, leads to
    \vspace*{-0.2cm}
    \begin{multline*}
        \begin{bmatrix}\star\end{bmatrix}^\top
        \begin{bmatrix}
            -P^{-1} & 0 \\ 0 & P^{-1}
        \end{bmatrix}    
        \begin{bmatrix}
            z_k \\
            (A+B_0k^\top)z_k + B_1 w_k + \varepsilon_k
        \end{bmatrix}
        \\
        + 
        \begin{bmatrix}\star\end{bmatrix}^\top
        \begin{bmatrix}
            Q_z & S_z \\ S_z^\top & R_z
        \end{bmatrix}
        \begin{bmatrix}
            w_k \\ k^\top z_k
        \end{bmatrix}
        - \tau^{-1}
        \begin{bmatrix}\star\end{bmatrix}^\top
        \begin{bmatrix}
            -L_\varepsilon^2 I & 0 \\
            0 & I
        \end{bmatrix}
        \begin{bmatrix}
            z_k \\ \varepsilon_k
        \end{bmatrix}
        < 0.
        \\[-1.25\baselineskip]
    \end{multline*}
    Further, we recall $u_k=k^\top z_k$, $w_k = z_k u_k$. Then, applying the S-procedure~\citep[cf.][]{boyd:vandenberghe:2004} yields
    $
        z_{k+1}^\top P^{-1} z_{k+1} - z_k^\top P^{-1} z_k < 0
    $
    for all $z_k\in\cZ$ and all $\varepsilon_k\in\cE$.
    Moreover, positive invariance of $\cZ_\mathrm{RoA}\subseteq\cZ$ holds for $k\geq 0$ due to similar arguments as in the proof of Theorem~\ref{thm:stability-condition-LFR-nominal}, and Assumption~\ref{ass:finite-gain} guarantees $\varepsilon_k\in\cE$ for all times.
    Thus, the controller robustly asymptotically stabilizes the LFR~\eqref{eq:LFR}, and, hence, system~\eqref{eq:dynamics-bilinear} for all $z_0\in\cZ_\mathrm{RoA}$. 
\vspace*{-0.15cm}
\end{pf}

Note that LMI~\eqref{eq:stability-condition-LFR} is of dimension ${(3N+1)\times (3N+1)}$ and contains $(N^2 + 3N)/2 + 1$ decision variables.

\begin{rem}
    Condition~\eqref{eq:stability-condition-LFR} of Theorem~\ref{thm:stability-condition-LFR} reduces to the nominal stabilization condition of Theorem~\ref{thm:stability-condition-LFR-nominal} for $L_\varepsilon=\tau = 0$, i.e., if there is no truncation error $\varepsilon_k$.
\vspace*{-0.05cm}
\end{rem}

As for the nominal case, we want to find a stabilizing controller for a large region $\cZ_\mathrm{RoA}\subseteq\cZ$ and, thus, maximize over $c$ after a successful design.

\begin{cor}\label{cor:stability-condition-nonlinear}
    Let $\Phi$ be injective and chosen such that Assumptions~\ref{ass:zero-if-zero},~\ref{ass:finite-gain} hold, and suppose $\Phi^{-1}$ exists locally and is continuous. Then, the controller resulting from Theorem~\ref{thm:stability-condition-LFR} asymptotically stabilizes the nonlinear system~\eqref{eq:dynamics-nonlinear} for all initial conditions $x_0$ satisfying $\Phi(x_0)\in\cZ_\mathrm{RoA}$.
\end{cor}
Local invertibility of $\Phi$ is satisfied if, e.g., the Jacobian of $\Phi$ has (uniformly) full rank.
The proof of Corollary~\ref{cor:stability-condition-nonlinear} follows standard Lyapunov arguments and is therefore omitted.

\vspace*{-0.125cm}
\section{Numerical example}\label{sec:numerical-example}
\vspace*{-0.19cm}
In this section, we illustrate our theoretical findings with the forced nonlinear Van der Pol oscillator
\vspace*{-0.04cm}
\begin{align*}
    \xdot_1 &= x_2, \\
    \xdot_2 &= \mu(1-x_1^2)x_2 - x_1 + u.
\vspace*{-0.06cm}
\end{align*}
The system is non-conservative with nonlinear damping determined by $\mu$ and a commonly studied example in nonlinear control. 
We choose $\mu=1$ for our simulations which are conducted in \textsc{Matlab} using YALMIP with the semidefinite programming solver MOSEK.

\begin{figure}[t]
    \centering
    \vspace*{-0.3\baselineskip}
    \begin{tikzpicture}[%
        /pgfplots/every axis x label/.style={at={(0.5,0)},yshift=-17.5pt},%
        /pgfplots/every axis y label/.style={at={(0,0.5)},xshift=-25pt,rotate=90},%
      ]%
		\begin{axis}[
			legend pos= south east,
			xlabel=$x_1$,
			xmin=-3,
			xmax=3,
			xtick distance=1,
            minor x tick num=4,
			ylabel=$x_2$,
			ymin=-3,
			ymax=3,
			ytick distance=1,
            minor y tick num=4,
            grid=both,
            width = 0.93\columnwidth,
            height = 0.65\columnwidth,
            minor grid style={gray!20},
            unbounded coords = jump
        ]
            \addplot[red,thick] table [x index=0,y index=1] {VdP_simulation_monomials.dat};
            \addplot[olive,thick,densely dashdotdotted] table [x index=2,y index=3] {VdP_simulation_monomials.dat};
            \addplot[cyan!60,thick,densely dashed] table [x index=4,y index=5] {VdP_simulation_delay.dat};
            \addplot[blue,thick,densely dotted] table [x index=4,y index=5] {VdP_simulation_monomials.dat};
            \addplot[only marks,mark=10-pointed star] coordinates {(1,-0.6)};
            \addplot[only marks,mark=+,thick] coordinates {(0,0)};
        \end{axis}
    \end{tikzpicture}
    \setbox1=\hbox{\begin{tikzpicture}[baseline]
        \draw[red,thick] (0,.6ex)--++(1.25em,0);
    \end{tikzpicture}}
    \setbox2=\hbox{\begin{tikzpicture}[baseline]
        \draw[olive,thick,densely dashdotdotted] (0,.6ex)--++(1.25em,0);
    \end{tikzpicture}}
    \setbox3=\hbox{\begin{tikzpicture}[baseline]
        \draw[cyan!60,thick,densely dashed] (0,.6ex)--++(1.25em,0);
    \end{tikzpicture}}
    \setbox4=\hbox{\begin{tikzpicture}[baseline]
        \draw[blue,thick,densely dotted] (0,.6ex)--++(1.25em,0); 
    \end{tikzpicture}}
    \setbox5=\hbox{\begin{pgfpicture}[baseline]
        \pgfuseplotmark{10-pointed star}
    \end{pgfpicture}}
    \setbox6=\hbox{\begin{pgfpicture}[baseline]
        \pgfuseplotmark{+}
    \end{pgfpicture}}
    \vspace*{-1.35\baselineskip}
    \caption{
        Open-loop trajectory (\usebox1) of the Van der Pol oscillator with initial condition (\usebox5), origin (\usebox6), and closed-loop trajectories using the controllers $u_\Psi$ (\usebox3), $u_\Phi$ (\usebox4), and, as comparison, the controller designed in~\cite{sinha:nandanoori:drgona:vrabie:2022} (\usebox2). 
    }
    \label{fig:controller-nominal}
    \vspace*{-0.25\baselineskip}
\end{figure}
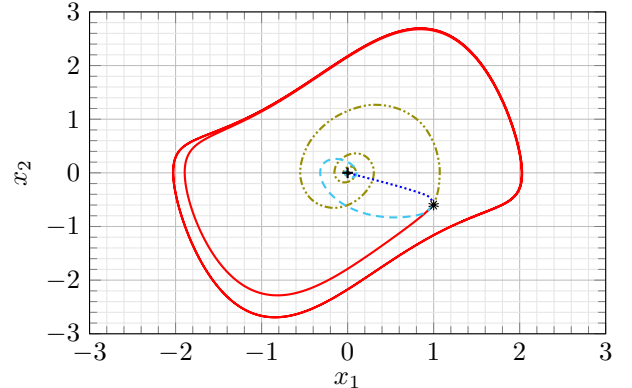
Our goal is to stabilize the origin by designing a control input $u$ based on data. 
To this end, we discretize the dynamics using the forward Euler method with step size $\tau=\SI{0.01}{s}$ and measure a trajectory of $\SI{20}{s}$.
We derive two different data-driven bilinear system representations for two lifting functions $\Psi$ and $\Phi$ as in Section~\ref{sec:bilinear-representation}: $\Psi$ is an $N_\Psi=32$-dimensional lifting consisting of delay coordinates with $d_x = 15,d_u=0$, and $\Phi$ contains all monomials up to degree 5 resulting in the dimension $N_\Phi = 20$. 
For the data generation, we use the initial condition $x_0^\mathrm{d}=\begin{bmatrix}-0.128 & -0.948\end{bmatrix}$ and a uniformly distributed input sequence with $u_k^\mathrm{d}\in[-1,1]$. 
The delay coordinates are constructed by simulating the unforced system for the first $d_x$ steps. 
Then, we design for each lifting the control laws $u_\Psi$ and $u_\Phi$, respectively, based on Theorem~\ref{thm:stability-condition-LFR-nominal}.

The corresponding LMI problems are both solved in less than one second. 
Further, $\Psi$ leads to an LMI of dimension $97\times 97$ with 561 decision variables, while $\Phi$ yields an LMI of dimension $61\times 61$ with 231 decision variables.
Fig.~\ref{fig:controller-nominal} illustrates the open-loop and the resulting closed-loop trajectories for the system with initial condition $x_0 = \begin{bmatrix}
    1 & -0.6
\end{bmatrix}^\top$. 
Both controllers are able to stabilize the origin of the nonlinear system, where, as shown, the controller $u_\Phi$ based on the monomial lifting achieves better behavior.
Moreover, we compare the resulting controllers to the one designed in~\cite{sinha:nandanoori:drgona:vrabie:2022}, where the same lifting function $\Phi$ was used. While the controller from the literature oscillates around the origin, our designed controller $u_\Phi$ leads to a direct trajectory to the origin.

For a better overview, we omit the closed-loop trajectory for a controller based on Theorem~\ref{thm:stability-condition-LFR}. 
Using the lifting~$\Phi$, condition~\eqref{eq:stability-condition-LFR} in Theorem~\ref{thm:stability-condition-LFR} is feasible for a finite-gain bound with constants $L_\varepsilon$ up to \num{1e-5}. Increasing $L_\varepsilon$ leads to a similar closed-loop behavior as before, but a more conservative controller and, thus, the trajectory shows more oscillations with slower convergence to the origin.

\vspace*{-0.1cm}
\section{Conclusion}\label{sec:conclusion}
\vspace*{-0.23cm}
In this paper, we presented a data-driven controller design method with stability guarantees for discrete-time control-affine nonlinear systems. 
We showed how measured data can be used to derive an LFR of the underlying system relying on Koopman operator theory. 
Next, we characterized the truncation error resulting from a finite-dimensional approximation via a finite-gain bound. 
We then used the LFR to obtain an LMI-based design procedure guaranteeing robust local stability for the bilinear system. 
While our data-driven controller design approach is of its own interest for bilinear systems, it also achieves closed-loop stability for nonlinear systems.
Finally, we demonstrated the proposed controller design method with the Van der Pol oscillator.

\vspace*{-0.2cm}
{\bibliography{literature-abbreviated}}

\end{document}